\documentclass[prl,aps,twocolumn,tightenlines,notitlepage,nofootinbib,preprintnumbers,letterpaper,superscriptaddress]{revtex4-1} 
\pdfoutput=1
\usepackage{epsfig}
\usepackage{amsfonts}
\usepackage{amsmath}
\usepackage{graphicx}
\usepackage{url}
\usepackage{color}
\usepackage{amssymb,amsmath,slashed,color,graphicx,bm}
\usepackage[T1]{fontenc}
\usepackage{relsize}
\usepackage{hyperref}
\hypersetup{allcolors=[rgb]{0.0 0.0 0.6},linkcolor=[rgb]{0.8 0.0 0.5}}

\hypersetup{colorlinks,citecolor= nicegreen,linkcolor= nicered}
\definecolor{nicered}{rgb}{0.7,0.1,0.1}
\definecolor{nicegreen}{rgb}{0.1,0.5,0.1}

\def\Fermilab{Theoretical Physics Department, Fermilab, P.O. Box 500, Batavia, IL 60510, USA}
\def\Northwestern{Department of Physics and Astronomy, Northwestern University, Evanston, IL 60208, USA}
\def\Berkeley{Department of Physics, University of California Berkeley, Berkeley, California 94720, USA}
\def\Loyola{Department of Physics, Loyola University Chicago, Chicago, IL 60660, USA}
\def\Carleton{Ottawa-Carleton Institute for Physics, Department of Physics, Carleton University, Ottawa, K1S 5B6, Canada}

\begin{document}

 \hfill{NUHEP-TH/19-15, FERMILAB-PUB-19-522-T}
\title{The Dodelson-Widrow Mechanism In the Presence of Self-Interacting Neutrinos}

\author{Andr\'e de Gouv\^ea}
\affiliation{\Northwestern}
\author{Manibrata Sen}
\affiliation{\Northwestern}
\affiliation{\Berkeley}
\author{Walter Tangarife}
\affiliation{\Loyola}
\author{Yue Zhang}
\affiliation{\Carleton}
\affiliation{\Northwestern}
\affiliation{\Fermilab}

\begin{abstract}
keV-scale gauge-singlet fermions, allowed to mix with the active neutrinos, are elegant dark matter (DM) candidates. They are produced in the early universe via the Dodelson-Widrow mechanism and can be detected as they decay very slowly, emitting X-rays. In the absence of new physics, this hypothesis is virtually ruled out by astrophysical observations. Here, we show that new interactions among the active neutrinos allow these sterile neutrinos to make up all the DM while safely evading all current experimental bounds. The existence of these new neutrino interactions  may manifest itself in next-generation experiments, including DUNE.  
\end{abstract}

\preprint{}

\maketitle 

A fourth neutrino $\nu_4$ with mass around the keV scale is an attractive dark matter (DM)
candidate. Other than its mass $m_4$, $\nu_4$ is characterized by a small active-neutrino component, parameterized by a mixing 
angle $\theta$. Even in the case where the new interaction eigenstate $\nu_s$ has no standard model (SM) quantum numbers and the mixing angle is tiny, nonzero values of $\theta$ allow for the nonthermal production of $\nu_4$  via neutrino oscillations in the early universe. Dodelson and Widrow showed that, with judicious choices of $m_4$ and $\theta$, $\nu_4$ can make up 100\% of the DM \cite{Dodelson:1993je}. The same Physics -- $\theta\neq 0$ -- allows the $\nu_4$ to decay, very slowly, into light neutrinos plus a photon~\cite{Pal:1981rm}. This renders this scenario falsifiable since one predicts the existence of an X-ray line from regions of the universe where DM accumulates. 

In more detail, $\nu_4$ is a linear combination of $\nu_s$ and $\nu_a$ ($a$ for active), the latter a linear combination of the standard model interaction-eigenstates $\nu_e,\nu_{\mu},\nu_{\tau}$. 
\begin{equation}
\nu_4 = \cos\theta\, \nu_s + \sin\theta \, \nu_a \ .
\end{equation}
We will be interested in the limit $\theta\ll 1$ and will refer to $\nu_4$ (and $\nu_s$) as the ``sterile'' neutrino whenever the usage of the term does not lead to any confusion. 

In a nutshell, the  Dodelson-Widrow (DW) mechanism works as follows. In the early universe, the active neutrinos are in thermal equilibrium with the other SM particles, whereas the sterile neutrino is out of equilibrium and assumed to have negligible initial abundance. The weak-interaction eigenstates $\nu_a$ are constantly produced and propagate freely in the plasma for a time interval $t$ before they are ``measured'' by another weak-interaction reaction. If this interval is long enough for 
neutrino oscillations to occur, the state $\nu(t)$ is no longer identical to its initial state and develops a $\nu_s$ component. When a ``measurement'' occurs, there is a small probability that the neutrino will collapse into a sterile state and, for the most part, remain in that state thereafter. This process occurs until the active neutrinos decouple from the rest of the universe and one is left with a relic population of sterile neutrinos.  

In the absence of new physics, this elegant mechanism is in tension with various current astrophysical observations.
A very light sterile-neutrino dark matter is inconsistent with various cosmological observations on small scales.
A conservative limit $m_4\gtrsim 2$~keV arises from phase space density derived for dwarf galaxies~\cite{Tremaine:1979we, Boyarsky:2008ju,Merle:2015vzu,Abazajian:2017tcc}.
This lower limit could be further improved by Lyman-$\alpha$ forest observations up to $\sim 30$\,keV~\cite{Yeche:2017upn}. Moreover, for $m_4\gtrsim 2$~keV, the mixing angle $\theta$ required so that $\nu_4$ makes up all of the DM leads to enough X-ray radiation from $\nu_4$ decays that it should have been observed by X-ray telescopes in the last decade~\cite{Watson:2011dw, Horiuchi:2013noa, Perez:2016tcq, Dessert:2018qih, Ng:2019gch}. On the other hand, the unidentified 3.5 keV photon line~\cite{Bulbul:2014sua, Boyarsky:2014jta} might be interpreted as evidence for decaying sterile-neutrino DM. This interpretation, however, favors  mixing angles that are small enough that the $\nu_4$ does not make up all of the DM. 

A popular new-physics solution to alleviate the tension highlighted above is to postulate the existence of a large lepton-number asymmetry in the universe~\cite{Shi:1998km}. This hypothesis is, in general, difficult to test unless the asymmetry is really large, while it does not explain why the lepton asymmetry is much larger than the baryon asymmetry. In this {\it letter}, we propose a new, experimentally testable sterile-neutrino-dark-matter 
production mechanism. We introduce a light scalar particle that mediates self-interactions among the active neutrinos. These new interactions enable the efficient production of sterile neutrinos in the early universe via the DW mechanism and allow one to resolve all the tensions in a straightforward way. The existence of the new interactions is testable; part of the allowed parameter space will be probed at future neutrino experiments, including DUNE \cite{Acciarri:2015uup}.

Concretely, we add the following interaction term involving SM neutrinos:
\begin{equation}\label{Lint}
\mathcal{L} 	\supset \frac{\lambda_\phi}{2} \nu_a \nu_a \phi + {\rm h.c.} \ ,
\end{equation}
where $\phi$ is a complex scalar with mass $m_{\phi}$ and we are only interested in the interactions with $\nu_a$, the linear combination of active neutrinos that mix with $\nu_s$. For the remainder of this letter, we only consider the effective two-neutrino $\nu_a-\nu_s$ system. SM gauge invariance of the new interaction can be restored with the insertion of the vacuum expectation value of the Higgs field. The operator can be further embedded in reasonable ultraviolet-complete models~\cite{Berryman:2018ogk, Kelly:2019wow,Blinov:2019gcj}.

The equation that describes the evolution of the sterile neutrino population as a function of time, for fixed neutrino energy $E\equiv x\,T$, where $T$ is the temperature of active neutrinos, is~\cite{Dodelson:1993je, Abazajian:2005gj, Hansen:2017rxr}
\begin{eqnarray}
\label{masterequation}
\frac{d f_{\nu_s}}{d z} & = & \frac{\Gamma \sin^22\theta_{\rm eff}}{4H z}  f_{\nu_a} \ , \\
\sin^22\theta_{\rm eff} & \simeq &  \frac{\Delta^2 \sin^22\theta}{\Delta^2 \sin^22\theta + \Gamma^2/4 + (\Delta \cos2\theta - V_T)^2} \ .
\end{eqnarray}
Here, $f_{\nu_s}(x, z)$ is the phase-space distribution function of the sterile neutrino, and we define the dimensionless evolution variable $z\equiv\mu/T$,  where $\mu \equiv 1\,$MeV. We restrict our discussions to $m_4\lesssim1$~MeV. $\Gamma$ is the total  interaction rate for the active neutrino, $\theta_{\rm eff}$ is the effective active-sterile neutrino mixing in the early universe, and $H$ is the Hubble rate. $f_{\nu_a}$ is the usual Fermi-Dirac thermal distribution function for the active neutrinos.
$\Delta \equiv m_4^2/(2E)$ is the neutrino oscillation frequency in vacuum, where $m_4\gg m_{1,2,3}$ and $V_T$ is the thermal potential experienced by the active neutrino. Note that Eq.~\eqref{masterequation} is valid as long as the number of relativistic degrees of freedom in the universe is unchanged, which is a good approximation as the dominant production occurs at temperatures below the QCD phase transition.

\begin{figure}[!t]
	\centerline{\includegraphics[width=0.3\textwidth]{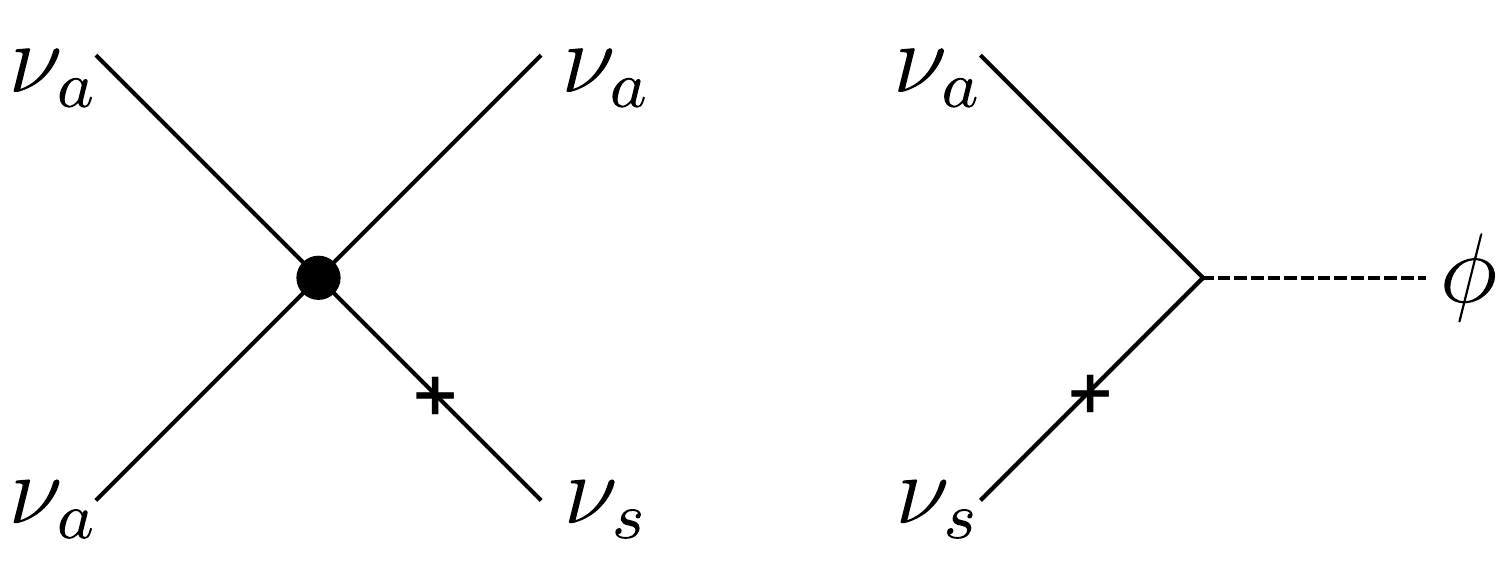}}
	\caption{Diagrams for sterile-neutrino production in the presence of the new neutrino interactions (Eq.(\ref{masterequation})), in the case of a heavy (left) or light (right) scalar mediator $\phi$.}\label{scatterdecay}
\end{figure}

The neutrino self-interaction mediated by $\phi$-exchange introduces new production channels for sterile neutrinos in presence of a nonzero $\theta$, as depicted in Fig.~\ref{scatterdecay}. This is reflected in Eq.~\eqref{masterequation} through contributions to the interaction rate $\Gamma$ and the thermal potential $V_T$. On the one hand, the contribution to $\Gamma$, in the very heavy $\phi$ limit ($m_\phi \gg T$), takes the form
\begin{equation}
\Gamma_\phi =\frac{7\pi \lambda^4 E_1 T^4}{864 m_\phi^4} \ .
\end{equation}
In contrast, a light $\phi$ ($T\gtrsim m_\phi$) can be directly produced in the plasma and neutrinos mainly self-interact through the decay and inverse-decay of $\phi$, with
\begin{equation}
\Gamma_\phi \simeq \frac{\lambda^2 m_\phi^2 T}{8\pi E_1^2} \left( \ln \left( 1 + e^{w} \right) - w\rule{0mm}{4mm}\right) \ ,
\end{equation}
where $w= m_\phi^2/(4E_1 T)$; see the Supplemental Material for details. The total rate $\Gamma$ is the sum of $\Gamma_\phi$, its charge conjugate, and the SM interaction rate, $\Gamma_{\rm SM}\sim G_{F}^{2} E T^{4}$~\cite{Abazajian:2005gj, Hansen:2017rxr}.

The thermal potential $V_T$ receives contributions from SM weak interactions, $V_T^{\rm SM} \sim G_F ET^4/M_W^2$~\cite{Abazajian:2005gj, Hansen:2017rxr}, where $G_F$ is the Fermi constant,
and from the new neutrino interaction. For generic mass $m_\phi$, the contribution from the new interaction is~\cite{Notzold:1987ik, Quimbay:1995jn}:
\begin{widetext}
\begin{equation}
\begin{split}
V_T^\phi(E, T) &= \frac{\lambda_\phi^2}{16\pi^2 E^2} \int_{0}^\infty dp \left[ \left( \frac{m_\phi^2 p}{2\omega} L_2^+(E, p) 
- \frac{4 E p^2}{\omega} \right)\frac{1}{e^{\omega/T}-1} + \left( \frac{m_\phi^2}{2} L_1^+(E, p) - 4 E p \right) \frac{1}{e^{p/T}+1} \right] \ , \\
L_1^+ (E, p) & = \ln \frac{4 p E + m_\phi^2}{4 p E - m_\phi^2}, \ \ \ \ \ 
L_2^+ (E, p)  = \ln \frac{\left( 2 p E + 2 E \omega + m_\phi^2 \right)\left(2 p E - 2 E \omega + m_\phi^2 \right)}
{\left( -2 p E + 2 E \omega + m_\phi^2 \right) \left( -2 p E - 2 E \omega + m_\phi^2 \right)} \ ,
\end{split}
\end{equation}
\end{widetext}
where $\omega=\sqrt{p^2 + m_\phi^2}$. This potential takes the asymptotic form
$V_T^\phi = - {7\pi^2 \lambda_\phi^2 E T^4}/({90 m_\phi^4})$ for $m_\phi \gg T$, and
$V_T^\phi = {\lambda_\phi^2 T^2}/({16 E})$ for $m_\phi \ll T$~\cite{Dasgupta:2013zpn,Jeong:2018yts}.

\begin{figure*}
	\centerline{\includegraphics[width=0.4\textwidth]{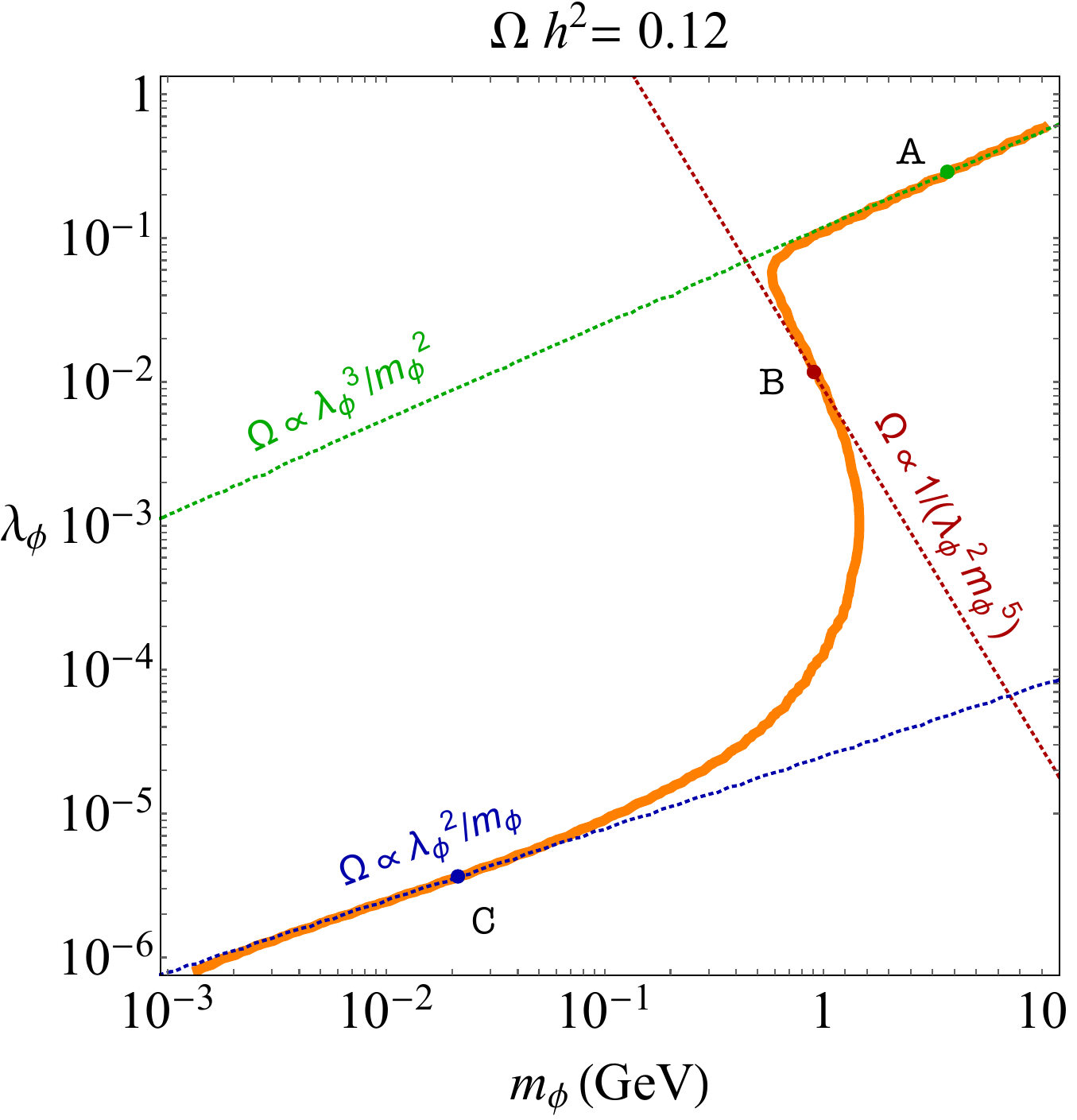}\qquad\qquad
	\includegraphics[width=0.405\textwidth]{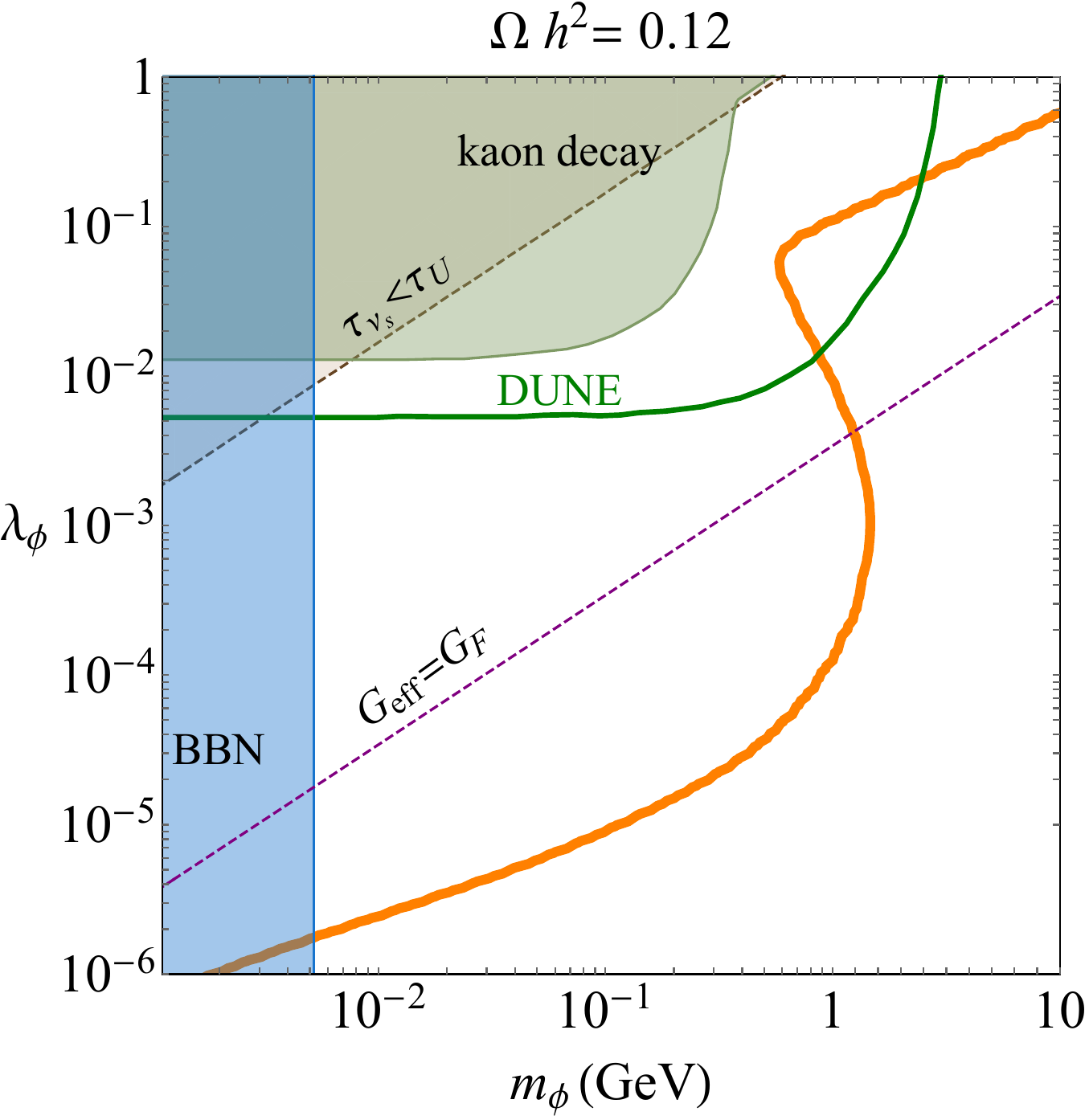}}
	\caption{Region in the $\lambda_\phi$ versus $m_\phi$ plane where the new neutrino interaction allows for the sterile neutrino to make up all the DM (orange curve). All other model parameters are fixed: $\sin^22\theta=7\times10^{-11}$ and $m_4=7.1\,$keV. In the left panel, the dotted lines represent the expected qualitative behavior of constant $\Omega$-contours when different approximations, discussed in the text, apply. 
In the right panel, the colored-shaded regions are excluded by imposing that the sterile neutrino DM candidate live longer than the age of our universe (brown), by  constraints from searches for rare charged kaon decays (green), and by BBN (blue). DUNE is sensitive, via the mono-neutrino channel, to the region above the green curve. The purple dashed curve indicates points where $G_{\rm eff}\equiv \lambda_\phi^2/m_\phi^2 = G_F$.}\label{RelicContours}
\end{figure*}

We numerically integrate Eq.~(\ref{masterequation}) up to $z\sim 10$. Note that, at this temperature, the relativistic approximation is no longer strictly valid for neutrinos heavier than $100\,{\rm keV}$; however these neutrinos are produced much earlier, and hence our calculation still holds. The yield $Y_{\nu_s}$ is given by the ratio $n_{\nu_s}/s$, where $s$ is the entropy density of the universe at $z=10$. The sterile neutrino relic density today can then be written as $\Omega = {Y_{\nu_s} s_0 m_4}/{\rho_0}$, where $s_0=2891.2\,{\rm cm^{-3}}$ is the entropy density today, and $\rho_0=1.05\times 10^{-5} h^{-2}\, {\rm GeV/cm^3}$ is the critical density. We identify the points in the parameter space where $\nu_4$ account for all of the DM. These are depicted in Fig.~\ref{RelicContours},  for fixed $m_4=7.1\,$keV, $\theta = 4\times 10^{-6}$, and $a=$~muon-flavor. Fig.~\ref{RelicContours} reveals that the DM-abundance constraint is satisfied along the ``S-shaped'' orange curve.

To understand the shape of the orange curve in Fig.~\ref{RelicContours}, we zoom into three specific points, labeled by {\tt A} (green), {\tt B} (red), {\tt C} (blue).  It is possible to derive the dependence of $\Omega$ on the model parameters by exploiting the behavior of the right-hand side of Eq.~(\ref{masterequation}) in some limiting cases. We define $z_0$ as the time when $\Delta \simeq |V_T|, \Gamma_a$, after which the effective mixing angle $\theta_{\rm eff}$ for $\nu_s$ production is no longer suppressed relative to the vacuum angle $\theta$. We also define $z_1\equiv \mu/m_\phi$ as the time when $\phi$ becomes heavy relative to the temperature of the universe.
For case {\tt A}, sterile neutrinos are mainly produced through the scattering of active neutrinos [Fig.~\ref{scatterdecay} (left)]. In this case, $df_{\nu_s}/dz \propto z^8$ for $z<z_0$, and $df_{\nu_s}/dz \propto z^{-4}$ for $z>z_0$, which implies $\nu_s$ is mainly produced around the time $z\sim z_0$. The resulting relic density is $\Omega \propto {\lambda_\phi^3\theta^2 m_4}/{m_\phi^2}$. This behavior is depicted by the dotted green line in Fig.~\ref{RelicContours} (left). For cases {\tt B} and {\tt C}, where $\lambda_\phi \ll 1$, $\nu_s$ is mainly produced through the decay of on-shell $\phi$ [Fig.~\ref{scatterdecay} (right)] while it is still light and well populated in the thermal plasma ($z<z_1$). One can estimate that $z_0 \sim 10^{-2} (\lambda_\phi/10^{-5}) ({\rm keV}/m_4)$. For case {\tt C}, $z_0<z_1$ and $\nu_s$ is dominantly produced during the epoch $z_0<z<z_1$, where $df_{\nu_s}/dz\propto z^{2}$. The resulting $\Omega \propto {\lambda_\phi^2 \theta^2}/{m_\phi}$ corresponds to the blue dotted line in Fig.~\ref{RelicContours} (left). In this case, the effective mixing angle  $\theta_{\rm eff}$ is close to the vacuum one, $\theta$.
In contrast, case {\tt B} has a relatively larger $\lambda_\phi$, thus $z_0>z_1$, and $\nu_s$ is mostly produced while the effective mixing angle is still suppressed due to thermal effects ($\theta_{\rm eff}\ll \theta$), leading to $df_{\nu_s}/dz \propto z^6$ during the epoch $z<z_1$. As a result, $\Omega \propto {m_4^4\theta^2}/({\lambda_\phi^2 m_\phi^5})$ corresponds to the red dotted line in Fig.~\ref{RelicContours} (left). 

\begin{figure}[ht]
	\centerline{\includegraphics[width=0.45\textwidth]{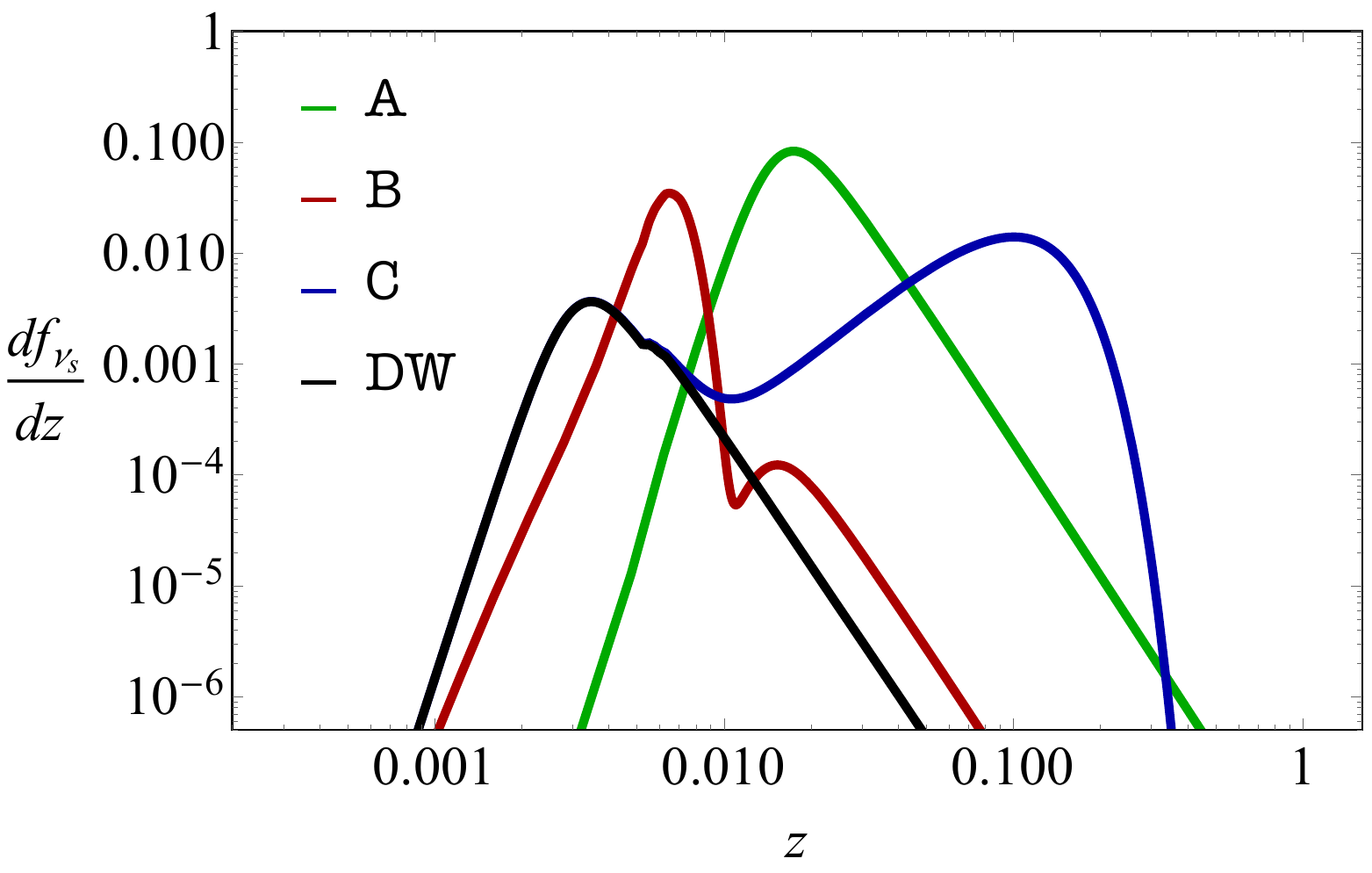}}
	\caption{Differential sterile neutrino production rate in the early universe, for the points {\tt A, B, C,} labeled in Fig.~\ref{RelicContours}. DW is the case of no neutrino interactions other than the ones in the SM.
	}\label{dfdz}
\end{figure}
The $z$ dependence of $df_{\nu_s}/dz$ for cases {\tt A, B, C,}  defined above, is depicted in Fig.~\ref{dfdz}. The values of  $\theta$ and $m_4$ are identical to the ones in  Fig.~\ref{RelicContours} and we concentrate on $x=1$ ($E=T$). Fig.~\ref{dfdz} allows one to identify the dominant sterile-neutrino-production epoch for each case. The black curve, labeled DW for the original Dodelson-Widrow scenario,  is the result obtained in the absence of new neutrino interactions. In cases {\tt A} and {\tt B}, the effective Fermi constant  $G_{\rm eff}= \lambda_\phi^2/m_\phi^2$ is much larger than $G_F$ and the new self-interaction at high temperatures enhances -- relative to the SM -- the thermal potential $V_T$, suppressing the effective mixing angle $\theta_{\rm eff}$. As a result, the onset of sterile-neutrino production in cases {\tt A, B} is delayed relative to that in the to the SM. Meanwhile, the new interaction is able to keep the active neutrino in thermal equilibrium for a longer period of time, relative to the SM case.\footnote{The extra peak around $z\sim0.007$ on the red curve (case {\tt B}) is due to an accidental resonant effect when $\Delta \cos2\theta = V_T$ in the denominator of Eq.~(\ref{masterequation}).}

Some of the new-physics parameter space can be explored in the laboratory. For example, Fig.~\ref{RelicContours} (right) depicts (green shaded region) the region of parameter space ruled out by searches for  $K^+\to \mu^+ \bar{\nu}_{\mu} + (\phi \to \nu\nu)$~\cite{Artamonov:2016wby}, assuming $a$ is the muon-flavor. If $m_\phi$ is below a few MeV (blue shade region in Fig.~\ref{RelicContours} (right)), $\phi$-production will significantly modify the expansion rate of the universe and affect the success of big-bang-nucleosynthesis (BBN)~\cite{Escudero:2019gvw}. 
Cosmological observations sensitive to the neutrino free-streaming length can be inferred from the CMB and translate into weaker bounds $(\lambda_\phi\gtrsim m_\phi/30\,{\rm MeV})$ on the parameter space~\cite{Berryman:2018ogk}.
A light mediator $\phi$, as discussed in~\cite{Berryman:2018ogk, Kelly:2019wow}, will be radiated during neutrino--matter interactions and will manifest itself as missing transverse momentum in fixed-target neutrino-scattering experiments. DUNE is expected to be sensitive to the parameter space above the thick green curve in Fig.~\ref{RelicContours} (right). Hence, for this value of $m_4$ and $\theta$, DUNE will be able to directly test some of the parameter space ($\lambda_\phi\gtrsim 10^{-2}$) where the sterile neutrinos account  for all of the DM. Other neutrino beam experiments sensitive to such $\phi$ emission have been discussed at length in \cite{Berryman:2018ogk}.

\begin{figure}[ht]
	\centerline{\includegraphics[width=0.45\textwidth]{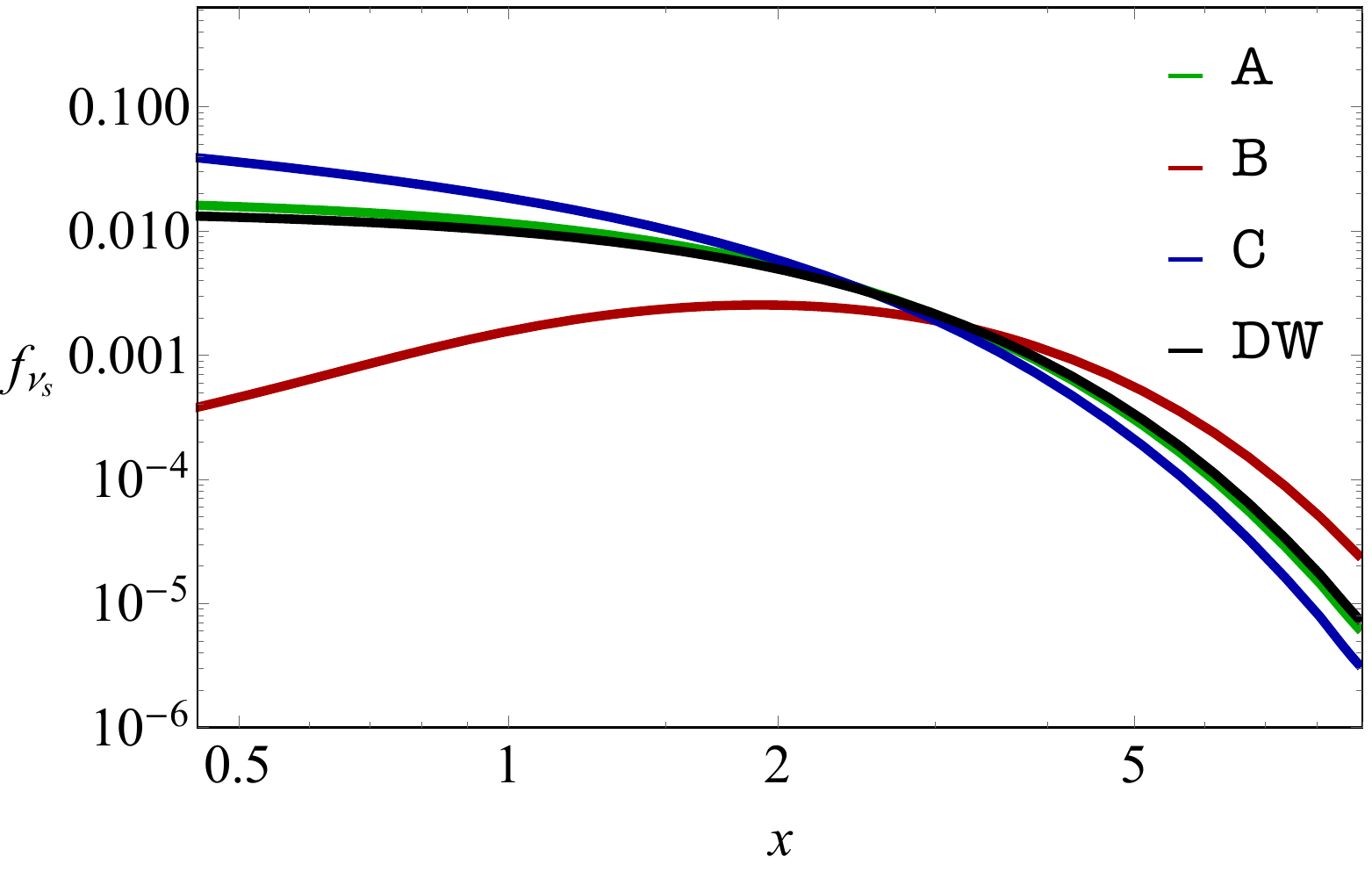}}
	\caption{Energy spectra of the sterile-neutrino dark-matter for cases {\tt A, B, C}, labeled in Fig.~\ref{RelicContours}. DW is the case of no neutrino interactions other than the ones in the SM.}\label{SterileDMSpectrum}
\end{figure} 
 Another imprint of the new neutrino interaction lies in the energy spectrum of the sterile-neutrino dark-matter, depicted in Fig.~\ref{SterileDMSpectrum} for cases {\tt A, B, C}. Case {\tt B} yields the hardest spectrum and hence the warmest DM while  in case {\tt C} the DM energy spectrum is the coolest. Different dark-matter spectra correspond to different free-streaming lengths in the early universe and may lead to identifiable features in small-scale structure observables such as the Lyman-$\alpha$ forest~\cite{Viel:2005qj, Boyarsky:2009ix}.

\begin{figure}[ht]
	\centerline{\includegraphics[width=0.48\textwidth]{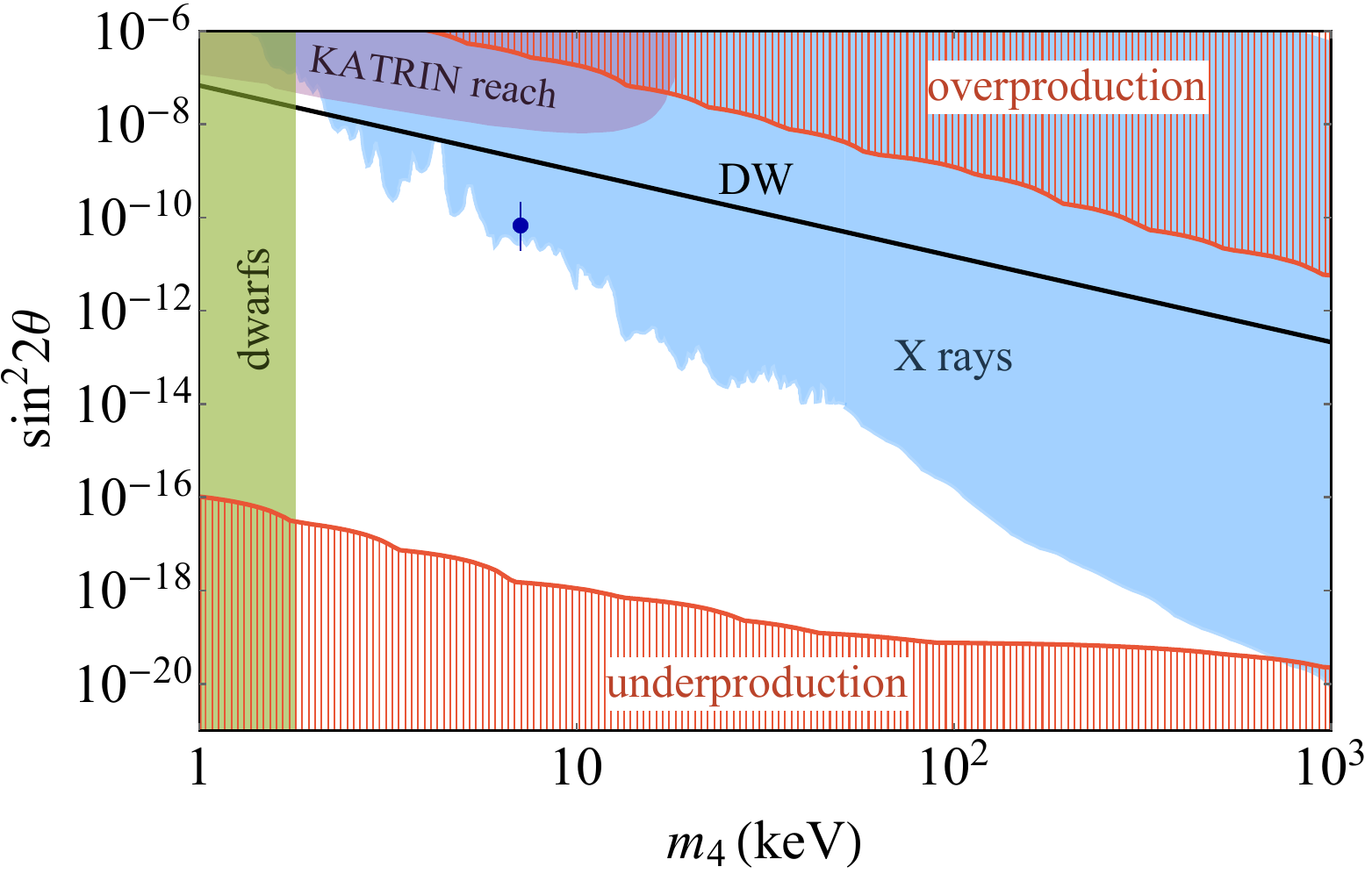}}
	\caption{Region of the $m_4$ versus  $\sin^22\theta$--parameter space where one can find values of $(\lambda_{\phi}$, $m_\phi)$ such that the sterile neutrino, produced via the DW mechanism, accounts for all the DM. In the absence of new neutrino interactions, one is confined to the black line. The solid, shaded regions of parameter space are excluded by X-ray (blue) and small-scale scale (green) observations, while the purple region indicates the expected sensitivity of KATRIN. In the hatched regions (red) there are no allowed values of $(\lambda_{\phi}$, $m_\phi)$ where the sterile neutrino makes up all the DM. The point with error bars corresponds to assigning the unidentified 3.5~keV X-ray line to DM sterile-neutrino decay \cite{Bulbul:2014sua}.}\label{SterileDMSearch}
\end{figure} 
Fig.~\ref{SterileDMSearch} depicts the region of the $\sin^22\theta$ versus $m_4$ parameter space where one can find values of $(\lambda_{\phi}$, $m_\phi)$ such that the sterile neutrino, produced via the DW mechanism, accounts for all the DM. The colored-shaded region has been excluded by searches for an excess of X-rays from DM rich-regions (blue)~\cite{Watson:2011dw, Horiuchi:2013noa, Perez:2016tcq, Dessert:2018qih, Ng:2019gch} and small-scale structure observations (green)~\cite{Tremaine:1979we, Boyarsky:2008ju}. The small-scale structure bounds are also consistent with the bounds arising out of free-streaming considerations of the sterile neutrinos~\cite{Abazajian:2001nj}.
The reach of the KATRIN experiment~\cite{Mertens:2015ila} is shaded in purple. The solid, black line labeled DW corresponds to the region of parameter space where sterile neutrinos account for all the DM in the absence of new neutrino interactions. It is, for the most part, in severe tension with existing astrophysical constraints. In contrast, when we ``turn on'' the new neutrino interactions in Eq.~(\ref{Lint}), the viable parameter space expands into the region between the two hatch-shaded regions (red). Some of the region above the DW line is now allowed, which is due to destructive interference between the SM and the new-interaction contributions to the thermal potential. After imposing the existing constraints from X-ray and dwarf galaxy observations, for any point in the empty (white) region, there is a family of values of $(\lambda_\phi, m_\phi)$ that lead the sterile neutrino to account for all the DM. The lower bound on $\theta$ as a function of $m_4$ is associated with the strong lower bound on $m_{\phi}$ from BBN, depicted in Fig.~\ref{RelicContours} (right). The keV sterile neutrino parameter space can also be strongly constrained from anomalous energy-loss arguments from SN1987A~\cite{Arguelles:2016uwb}. 

Fig.~\ref{SterileDMSearch} reveals that the presence of the new neutrino interaction extends the allowed parameter from a narrow line to a broad band. Some of the extended parameter space will be challenged by the next generations of X-ray observations \cite{Adams:2019nbz}, but we expect some of the parameter space associated to the smallest mixing angles will remain available even in the absence of a discovery from these astrophysical probes. On the other hand, for those values of $\theta$, one requires new neutrino interactions that are stronger than what is depicted in Fig.~\ref{RelicContours} and hence searches at future laboratory experiments, including DUNE, are expected to play a nontrivial role. 

It is amusing that a new interaction that acts exclusively on the active neutrinos can enhance the production of sterile neutrinos in the early universe via the DW mechanism. While here we concentrated on the scalar interaction Eq.~(\ref{masterequation}), we speculate that similar results would arise from different types of neutrino self-interactions, including those mediated by a new gauge boson which couples to the active neutrinos. These are associated to slightly different phenomenological and model-building challenges.  We also note that the impact of new interactions involving the sterile neutrino to the dynamics of these light fermions in the early universe -- different from what we are discussing here -- have been explored in \cite{Babu:1991at,Dasgupta:2013zpn,Hannestad:2013ana,Shuve:2014doa,Mirizzi:2014ama,Cherry:2016jol,Chu:2018gxk,Johns:2019cwc}.

\emph{Acknowledgments\,}--We would like to thank Kevin Kelly for useful discussions, and Daniel Aloni for correspondence and pointing out a missing factor in one of the equations in a previous version.  The work of AdG was supported in part by DOE grant \#DE-SC0010143. MS acknowledges support from the National Science Foundation, Grant PHY-1630782, and to the Heising-Simons Foundation, Grant 2017-228. The research of WT is supported by the College of Arts and Sciences of Loyola University Chicago. The work of YZ is supported by the Arthur B. McDonald Canadian Astroparticle Physics Research Institute.

\appendix
\section*{Supplemental Material}
\subsection{Calculation of $\Gamma_\phi$ for on-shell production of $\phi$}
The neutrino self-interaction mediated by $\phi$ introduces a new production channel for sterile neutrinos. This is governed by the scattering rate $\Gamma_\phi$, see Eq.\,(5) and Eq.\,(6). The cross section for the process $\nu_a \nu_a \leftrightarrow \nu_a \nu_a $ is 
\begin{equation}
\sigma\,(\nu_a \nu_a \leftrightarrow \nu_a \nu_a ) = \frac{\lambda_\phi^4 s}{32\pi \left(\left(s-m_\phi^2\right)^2+  m_\phi^2 \gamma_\phi^2 \right)}\,.
\end{equation}
Here $\gamma_\phi= \lambda_\phi^2 m_\phi/ (32\pi)$ is the decay rate of $\phi$, and $s=2 E_{\rm in} E_{\rm tar } (1-{\rm cos}\,\theta)$ for neutrinos scattering with energy $E$ and angle $\theta$. To calculate $\Gamma_\phi$, we have to average over the thermal distribution of the target $\nu_a$ for a fixed energy of the incoming $\nu_a$, i.e.,
\begin{equation}
\Gamma_\phi = \int\,\frac{d^3\,p_{\rm tar}}{(2\pi^3)}\frac{1}{{\rm e}^{E_{\rm tar}/T}+1}\,\sigma(\nu_a \nu_a \leftrightarrow \nu_a \nu_a)\,v_\text{Møller} \, ,
\end{equation}
where the Møller velocity between the incoming and the target particle is $v_\text{Møller}  =\sqrt{\left(  \vec{v}_{\rm in} - \vec{v}_{\rm tar}\right)^2 - (\vec{v}_{\rm in} \times \vec{v}_{\rm tar})^2 }$.
If the $\phi$ is produced on-shell, we use the narrow-width approximation to write 
\begin{equation}
\sigma(\nu_a \nu_a \rightarrow \phi \rightarrow \nu_a \nu_a)  = \frac{ \lambda_\phi^4 s}{32\pi }\frac{\pi}{m_\phi \gamma_\phi}\delta (s - m_\phi^2)\,.
\end{equation}
This yields
\begin{equation}
\Gamma_\phi \simeq  \frac{\lambda^2 m_\phi^2 T}{8\pi E_{\rm in}^2} \left( \ln \left( 1 + e^{w} \right) - w \rule{0mm}{4mm}\right)\ ,
\end{equation}
where $w= m_\phi^2/(4E_{\rm in} T)$.

\end{document}